\documentclass{jfm}

\usepackage{graphicx}
\usepackage{epstopdf}
\usepackage{amsmath}
\usepackage{amsfonts}
\usepackage{amssymb}
\usepackage{mathrsfs}
\usepackage{subfigure}
\usepackage{natbib}
\usepackage{bm}
\usepackage{color}
\usepackage{units}
\usepackage[dvipsnames]{xcolor}

\ifCUPmtlplainloaded \else
  \checkfont{eurm10}
  \iffontfound
    \IfFileExists{upmath.sty}
      {\typeout{^^JFound AMS Euler Roman fonts on the system,
                   using the 'upmath' package.^^J}%
       \usepackage{upmath}}
      {\typeout{^^JFound AMS Euler Roman fonts on the system, but you
                   dont seem to have the}%
       \typeout{'upmath' package installed. JFM.cls can take advantage
                 of these fonts,^^Jif you use 'upmath' package.^^J}%
      }
  \else
  \fi
\fi

\ifCUPmtlplainloaded \else
  \checkfont{msam10}
  \iffontfound
    \IfFileExists{amssymb.sty}
      {\typeout{^^JFound AMS Symbol fonts on the system, using the
                'amssymb' package.^^J}%
       \usepackage{amssymb}%

      }{}
  \fi
\fi

\ifCUPmtlplainloaded \else
  \IfFileExists{amsbsy.sty}
    {\typeout{^^JFound the 'amsbsy' package on the system, using it.^^J}%
     \usepackage{amsbsy}}
    {\providecommand\boldsymbol[1]{\mbox{\boldmath $##1$}}}
\fi

\newcommand{\m}{\mathbf{m}}
\newcommand{\w}{\omega}
\newcommand{\s}{\mathbf{s}}
\newcommand{\uu}{\mathbf{u}}
\newcommand{\Aa}{\mathbf{A}}

\usepackage{psfrag,graphicx,epsfig,color}

\newlength{\SGfigurewidth}
\newlength{\SGlabelwidth}

\newcommand{\tr}{\mathrm{tr}}

\newcommand{\RNum}[1]{\uppercase\expandafter{\romannumeral #1\relax}}

\title[velocity gradient in homogeneous compressible turbulence]
{Low-order moments of the velocity gradient in homogeneous compressible  turbulence}

\author[Yang, Fang, Fang \etal]
{P.-F. Yang$^{1,2,3}$, J. Fang$^{4}$, L. Fang$^{5}$, A. Pumir$^{6,7}$, and H. Xu$^{3}$}

\affiliation{
$^1$  The State Key Laboratory of Nonlinear Mechanics, Institute of Mechanics, Chinese Academy of Sciences, Beijing 100190, China
\\[\affilskip]
$^2$   School of Engineering Sciences, University of Chinese Academy of Sciences, Beijing 100049, China
\\[\affilskip]
$^3$ Center for Combustion Energy and School of Aerospace Engineering\\Tsinghua University, Beijing, 100084, China
\\[\affilskip]
$^4$ Scientific Computing Department, STFC Daresbury Laboratory, Warrington WA4 4AD, UK
\\[\affilskip]
$^5$ LCS, Ecole Centrale de P\'ekin, Beihang University, Beijing 100191, China
\\[\affilskip]
$^6$ Laboratoire de Physique, Ecole Normale Sup\'erieure de Lyon, CNRS\\Universit\'e de Lyon, Lyon,  F-69007 France
\\[\affilskip]
$^7$ Max Planck Institute for Dynamics and Self-Organization, G\"ottingen, D-37077, Germany
}

\pubyear{}
\volume{}
\pagerange{}
\date{\today}
\begin{document}
\maketitle

\begin{abstract}
We derive from first principles analytic relations for the second and third order moments of the velocity gradient $m_{ij} = \frac{\partial u_i}{\partial x_j}$ in compressible turbulence, which generalize known relations in incompressible flows.
These relations, although derived for homogeneous flows, hold 
approximately for a mixing layer. 
We also discuss how to apply these relations to determine all the second and third moments of the 
velocity gradient experimentally.
\end{abstract}

\maketitle

\section{Introduction}
\label{sec:intro}

In high Reynolds number flows, the velocity field, $\uu$, develops
very sharp gradients~\citep{Frisch95,Sreeni97}, resulting in extremely large fluctuations of
the velocity gradient $\m$, or $m_{ij} = \frac{\partial u_i}{\partial x_j}$.
For this reason, an accurate description of
the velocity gradient is essential to understand 
the small-scale properties of turbulence~\citep{Meneveau2011ANNU}. 
In the case of incompressible turbulence, much emphasis has been put on enstrophy, defined as $\frac{1}{2} \w_i \w_i$, where ${\bm \omega} = \nabla \times \uu$. Its amplification rate, 
known as vortex stretching, $\w_i s_{ij} \w_j$, where $\s$ is the symmetric part of $\m$, or the rate of
strain tensor, have
been thoroughly studied to investigate the production of small scales in the 
flow~\citep{Tsinober2009,Buaria20a}.
It should be kept in mind that a thorough description of small scales involves the full tensor $\m$, and not just vorticity~\citep{Meneveau2011ANNU}.

Two remarkable constraints on the second and third moments of $\m$ have been established by \cite{Betchov-1956} for homogeneous and 
incompressible flows:
\begin{equation}
\langle \tr(\m^2 )\rangle = 0 \quad {\rm and} \quad \langle \tr(\m^3) \rangle = 0,
\label{eq:Betchov}
\end{equation}
where $\langle \, \rangle$ denotes ensemble average. 
In addition, when the flow is isotropic, the identities \eqref{eq:Betchov} 
lead to remarkable simplifications, allowing
the second and third order moments of the velocity gradient 
tensors $\Aa^{(2)} = \langle \m^2 \rangle$ and $\Aa^{(3)} = \langle \m^3 \rangle $
to be expressed in terms of only one scalar quantity \citep{Pope2000}.
This implies that in an isotropic flow, $\Aa^{(2)}$ and $\Aa^{(3)}$ can be completely determined from the measurement of only one component of the velocity gradient tensor, e.g., using hot-wire probes. Although strictly zero in homogeneous incompressible flows, the values of 
$\langle \tr(\m^2 )\rangle$ and $\langle \tr(\m^3) \rangle $ remain very small even when spatial inhomogeneity is strong, such as in channel 
flows~\citep{Bradshaw1993,PumirXu-2016}. Effectively, this can be understood as 
a consequence of the slow variation of the mean flow properties, compared
to the very fast variation of turbulence at small scales. Nonetheless,
the structure of the velocity gradient tensors, and in particular of $\Aa^{(2)}$ 
and $\Aa^{(3)}$, strongly deviate from the isotropic
case \citep{Bradshaw1993,Vreman14,pumir2017structure}.

Here, we focus on the tensors $\Aa^{(2)}$ and $\Aa^{(3)}$
in compressible flows. 
How compressibility affects the structure of the velocity gradient tensor
has been studied in homogeneous isotropic 
flows \citep{Pirozzoli2004isoCompressible, Wang2012compressiso,Fang2016Fourthorder,Wang2018shockwaves}, 
in homogeneous shear flows \citep{Ma2016compressibleshear, ChenWang2019compressshear}, 
in mixing layers \citep{Vaghel2015} 
and in boundary layers \citep{Chu2014compressibleBL}. 
The dynamics of the velocity gradient tensors in compressible flows have been studied by \cite{Suman2009,Suman2011,Suman2013}
starting from the homogenized Euler equation.

The purpose of this work is to generalize the Betchov relations,
Eq.~\eqref{eq:Betchov}, to homogeneous compressible flows, which allows us to establish the general structure
of $\Aa^{(2)}$ and $\Aa^{(3)}$ similar to the case of incompressible flows. We validate these relations with direct numerical simulations (DNS) of homogenous isotropic compressible turbulence, and demonstrate that they still approximately hold in flows with strong inhomogeneity, i.e., in a turbulent mixing layer. Whereas only one 
scalar was sufficient to 
capture the full structure
of $\Aa^{(2)}$ or $\Aa^{(3)}$ in incompressible flows, for compressible turbulence, 2 independent parameters for $\Aa^{(2)}$ and 4 parameters for $\Aa^{(3)}$ are required.
Nonetheless, as we show, these
relations can be used to construct the full tensors of $\Aa^{(2)}$ and $\Aa^{(3)}$ from stereo-PIV measurements in compressible flows.

\section{Second- and third- order relations of compressible turbulence}
\label{sec:2nd3rdcompressible}

Following \cite{Betchov-1956}, we first consider the second-order moment $\Aa^{(2)}$:
\begin{equation}
A^{(2)}_{ijji} =  \left \langle  \overline{\mathbf{m}^2}  \right\rangle  = \left \langle \frac{\partial u_i}{\partial x_j} \frac{\partial u_j}{\partial x_i} \right\rangle = \left \langle \frac{\partial u_i}{\partial x_i} \frac{\partial u_j}{\partial x_j} \right\rangle = \left \langle  \overline{\mathbf{m}}^2  \right \rangle = A^{(2)}_{iijj},
\label{eq:Betchov2compressible}
\end{equation}
in which we used the notation $\overline{\mathbf{X}} = \tr(\mathbf{X})$ and the summation convention, and assumed homogeneity.
For the third moments, we notice a general relation between the traces of third-order moments involving the gradients of 3 homogeneous vector fields, 
Eq.~\ref{eq:third_order_mixrelations} in the Appendix, which yields,
for the special case where the 3 fields are identical,
\begin{equation}
A^{(3)}_{ijjkki} = \left \langle  \overline{\mathbf{m}^3}  \right\rangle  
= \frac{3}{2} \left \langle \overline{\mathbf{m}}\,\overline{\mathbf{m}^2}\right \rangle - \frac{1}{2} \left \langle  \overline{\mathbf{m}}^3  \right \rangle
= \frac{3}{2} A^{(3)}_{iijkkj} - \frac{1}{2} A^{(3)}_{iijjkk}.
\label{eq:Betchov3compressible}
\end{equation}
A straightforward consequence of Eqs.~\eqref{eq:Betchov2compressible} and \eqref{eq:Betchov3compressible} can
be expressed by introducing 
$\s \equiv (\m + \m^T)/2 - (\overline{\m}/3) \mathbf{I}$ and
$\mathbf{w} \equiv (\m - \m^T)/2$:
\begin{equation}
\left \langle  \overline{\mathbf{s}^2}  \right \rangle = - \left \langle  \overline{\mathbf{w}^2}  \right \rangle + \frac{2}{3} \left \langle  \overline{\mathbf{m}}^2  \right \rangle \, ,
\label{eq:Betchov2compressibleB}
\end{equation}
\begin{equation}
\left \langle  \overline{\mathbf{s}^3}  \right \rangle = - 3 \left \langle  \overline{\mathbf{wsw}}  \right \rangle + \frac{1}{2} \left \langle  \overline{\mathbf{m}} \overline{\mathbf{m}^2} \right \rangle - \frac{5}{18} \left \langle  \overline{\mathbf{m}}^3 \right \rangle \, .
\label{eq:Betchov3compressibleB}
\end{equation}
Equations formally similar to Eqs.~\eqref{eq:Betchov2compressible}-\eqref{eq:Betchov3compressibleB} were derived by \cite{Yang-2020} for the perceived velocity gradient tensor based on regular tetrahedra in incompressible homogeneous turbulence (see their Eqs. (3.11), (3.12), and (3.37), (3.38)).

In the restricted case of isotropic flows,
$\Aa^{(2)}$ is expressible as \citep{Pope2000}:
\begin{equation}
A^{(2)}_{ijkl} = \alpha \delta_{ij} \delta_{kl} + \beta \delta_{ik} \delta_{jl} + \gamma \delta_{il} \delta_{jk}.
\label{eq:Aijkl_iso}
\end{equation}
With this notation, $\langle \overline {\m^2} \rangle = 9 \alpha + 3 \beta + 3 \gamma$ and $\langle \overline{\m}^2 \rangle = 3 \alpha + 3 \beta + 9 \gamma$, so Eq.~\eqref{eq:Betchov2compressible} implies that $\alpha = \gamma$,
which means that only 2 quantities are necessary to fully determine the 
second order tensor $\Aa^{(2)}$. These can be determined in an 
experiment measuring $\frac{\partial u_1}{\partial x_1}$ and $\frac{\partial u_1}{\partial x_2}$ by e.g., planar PIV, or 2-component laser doppler velocimetry (2C LDV) or hot-wire anemometry
with cross-wires using Taylor's frozen turbulence hypothesis, and noticing that
$A^{(2)}_{1111} = \alpha + \beta + \gamma$, $A^{(2)}_{1122} = \alpha$, and $A^{(2)}_{1212} = \beta$. 
Interestingly, $\alpha = \gamma$ implies that:
\begin{equation}
\left \langle \frac{\partial u_1}{\partial x_1}\frac{\partial u_2}{\partial x_2}\right \rangle = A^{(2)}_{1122} = A^{(2)}_{1221} = \left \langle \frac{\partial u_1}{\partial x_2}\frac{\partial u_2}{\partial x_1}\right \rangle.
\label{eq:Aijkl_component_identity}
\end{equation}

Additionally, we notice that $ \langle \overline{\mathbf{m}}^2 \rangle = 9 \alpha + 3 \beta + 3 \gamma = 3(4\alpha+\beta) \geqslant 0$, 
thus $4\alpha + \beta \geqslant 0$, and $\langle  \overline{\mathbf{w}^2}  \rangle = \frac{1}{2} ( \langle  \overline{\mathbf{m}}^2 \rangle - \langle \overline{\mathbf{m}\mathbf{m^T}}\rangle) =3(\alpha - \beta) \leqslant 0$, thus $\alpha \leqslant \beta$.
These two inequalities constrain the the ratio of components of $\Aa^{(2)}$:
\begin{equation}
\frac{1}{3} = \frac{\beta}{2\beta + \beta} \leqslant \frac{A^{(2)}_{1212}}{A^{(2)}_{1111}} = \frac{\beta}{2\alpha+\beta} = 2 \times \frac{\beta}{(4\alpha + \beta)+\beta} \leqslant 2 .
\label{eq:Aijkl_component_inequality1}
\end{equation}
The two limiting equalities arise when $(4\alpha + \beta ) = 0$, 
i.e., when the flow is incompressible, or when $\alpha = \beta$, i.e., when the flow is irrotational.

In isotropic flows, the third-order tensor $\Aa^{(3)}$
can be expressed as \citep{Pope2000}:
\begin{align}
& A^{(3)}_{ipjqkr}  =  a_1 \delta_{ip} \delta_{jq} \delta_{kr} + a_2 ( \delta_{ip} \delta_{jk} \delta_{qr} +  \delta_{jq} \delta_{ik} \delta_{pr} + \delta_{kr} \delta_{ij} \delta_{pq}) \notag \\
& +  a_3 ( \delta_{ip} \delta_{jr} \delta_{qk} + \delta_{jq} \delta_{ir} \delta_{pk} + \delta_{kr} \delta_{iq} \delta_{pj})+ a_4 (\delta_{iq} \delta_{pk} \delta_{jr} + \delta_{ir} \delta_{pj} \delta_{qk}) \notag \\
& + a_5 (\delta_{ij} \delta_{pk} \delta_{qr} + \delta_{ij} \delta_{qk} \delta_{pr} + \delta_{ik} \delta_{pj} \delta_{qr} + \delta_{ik} \delta_{rj} \delta_{pq} + \delta_{jk} \delta_{qi} \delta_{pr} + \delta_{jk} \delta_{ri} \delta_{pq}),
\label{eq:Aipjqkr_iso}
\end{align}
from which it is easy to obtain following expressions for the invariants of $\mathbf{m}$:
\begin{align}
& \left \langle \overline{\mathbf{m}}^3 \right \rangle = 27 a_1 + 27 a_2 + 27 a_3 + 6 a_4 + 18 a_5,
\label{eq:A_iso_invariants_1} \\
& \left \langle \overline{\mathbf{m}^3} \right \rangle = 3 a_1 + 9 a_2 + 27 a_3 + 30 a_4 + 36 a_5,
\label{eq:A_iso_invariants_2} \\
& \left \langle \overline{\mathbf{m}}~\overline{\mathbf{m}^2} \right \rangle = 9 a_1 + 15 a_2 + 33 a_3 + 18 a_4 + 30 a_5,
\label{eq:A_iso_invariants_3} \\
& \left \langle \overline{\mathbf{m}}~\overline{\mathbf{m}\mathbf{m^T}} \right \rangle = 9 a_1 + 33 a_2 + 15 a_3 + 6 a_4 + 42 a_5,
\label{eq:A_iso_invariants_4} \\
& \left \langle \overline{\mathbf{m}^2\mathbf{m^T}} \right \rangle = 3 a_1 + 21 a_2 + 15 a_3 + 12 a_4 + 54 a_5 .
\label{eq:A_iso_invariants_5}
\end{align}
In the incompressible case, with $\overline{\m} = 0$ and $\langle \overline{\m^3 } \rangle = 0$~\citep{Betchov-1956}, the left-hand sides of 
Eqs.~\eqref{eq:A_iso_invariants_1}-\eqref{eq:A_iso_invariants_4} are all $0$,
which provides 4 constraints to express $a_1, \ldots, a_5$ in terms
of one scalar quantity. 
For compressible turbulence, only one relation 
is obtained by plugging
Eqs.~\eqref{eq:A_iso_invariants_1} - \eqref{eq:A_iso_invariants_3} 
into Eq.~\eqref{eq:Betchov3compressible}, which leads to $a_1 = 3 a_3 - 2 a_4$. 
Thus 4 independent constants are needed to completely determine $\Aa^{(3)}$. Their experimental determination
would require techniques like stereo-PIV or 3C LDV with frozen turbulence hypothesis, giving access to spatial derivatives of the third velocity component normal to the plane of imaging, e.g., $A^{(3)}_{113232}$. 
With these components, and using Eq.~\eqref{eq:Aipjqkr_iso}, one can determine 4 independent constants, say, $a_2, \ldots, a_5$, as
\begin{align}
& a_2 = A^{(3)}_{113232},
\label{eq:A_iso_as_1} \\
& a_3 = \frac{1}{6} A^{(3)}_{111111} - \frac{1}{2} A^{(3)}_{212111},
\label{eq:A_iso_as_2} \\
& a_4 = \frac{1}{3} A^{(3)}_{111111} -  A^{(3)}_{212111} - \frac{1}{2} A^{(3)}_{111122} + \frac{1}{2} A^{(3)}_{113232},
\label{eq:A_iso_as_3} \\
& a_5 = \frac{1}{2} A^{(3)}_{212111} -  \frac{1}{2} A^{(3)}_{113232}.
\label{eq:A_iso_as_4}
\end{align}
The other invariants of $\m$, e.g.,
$\left \langle \overline{\mathbf{s}^3} \right \rangle$, $\left \langle \overline{\mathbf{wsw}}\right \rangle$, $\left \langle \overline{\mathbf{m}}~\overline{\mathbf{s^2}} \right \rangle$, and $\left \langle \overline{\mathbf{m}}~\overline{\mathbf{w^2}} \right \rangle$, can also be represented by these constants. In particular, we note that
\begin{equation}
\left \langle \overline{\mathbf{m}}~\overline{\mathbf{w^2}} \right \rangle = -9 a_2 + 9 a_3 + 6 a_4 -6 a_5 .
\label{eq:A_iso_invariants_9}
\end{equation}

\section{DNS of compressible turbulence}
\label{sec:DNS}

To test the relations presented above in various turbulent flow configurations, 
we numerically solved the three-dimensional compressible Navier-Stokes equations:
\begin{equation}
\begin{aligned}
& \frac{\partial \rho}{\partial t}+\frac{\partial \rho u_i}{\partial x_i}=0, \\
& \frac{\partial \rho u_i}{\partial t}+\frac{\partial }{\partial x_j}\left( \rho u_i u_j + P\delta_{ij}\right)-\frac{\partial}{\partial x_j}\sigma_{ij}=0, \\
& \frac{\partial E}{\partial t}+\frac{\partial }{\partial x_j} \left [(E + P) u_j  \right ]-\frac{\partial}{\partial x_j}\left (\sigma_{ij} u_i-Q_j  \right )=0,
\end{aligned}
\label{eq:nse}
\end{equation}
in which $\rho$ is the fluid density, $P$ is the pressure,
$E=\frac{1}{2}\rho u_iu_i+P/(\gamma -1)$ is the total energy with $\gamma=1.4$ being the ratio of specific heats,
$\sigma_{ij}=\mu \left ( \frac{\partial u_i }{\partial x_j} + \frac{\partial u_j }{\partial x_i} - \frac{2}{3}\frac{\partial u_k }{\partial x_k} \delta_{ij} \right )$ is the viscous stress tensor with the effect of bulk viscosity neglected~\citep{Pan17}, and
$Q_j=\kappa \frac{\partial T}{\partial x_j}$ is the heat flux,
where $\mu$ is the viscosity coefficient determined by the temperature, $T$, via Sutherland’s law, and $\kappa$ is the thermal conductivity. The temperature is related to fluid pressure and density via the ideal gas law $P=\rho R T$ and the gas constant $R$. 
The set of equations ~\eqref{eq:nse} are solved with high-order finite difference method.
Namely, the convection terms are computed by a seventh-order low-dissipative monotonicity-preserving scheme \citep{FangLi886} 
in order to capture shock-waves in a compressible flow while preserving the capability of resolving 
small-scale turbulent structures.
The diffusion terms are computed by a sixth-order compact central scheme \citep{Lele429} with a domain decoupling scheme for parallel computation \citep{FangGao1850}.
The time-integration is performed by a three-step third-order total 
variation diminishing Runge-Kutta method~\citep{GottliebShu522}.
The flow solver used in the present study is ASTR, an in-house code 
previously tested in DNS of various compressible turbulent flows with and without shock-waves \citep{FangLi886,FangYao1856,FangYao977,Fang2009}.

We first discuss the results of decaying isotropic compressible turbulence. The computational domain is a $(2\pi)^3$ cube with periodic boundary conditions in all three directions.
The initial flow is a divergence-free random velocity field with a sharply decaying spectrum:
$E( k ) =A k^4e^{-2k^2/k_{0}^{2} }$,
which peaks at $k = k_0 = 4$, and the value of $A$ determines the kinetic energy at 
time $t=0$.
The density, pressure and temperature are all initialized to constant values, 
similar to the ``IC4'' run of \cite{Samtaney417}.
The initial turbulent Mach number based on the root-mean-square velocity $u'=\sqrt{{u_i}{u_i}}$ is $Ma_t= u' / \left\langle{c}\right\rangle $=2.0, where $c= \sqrt{\gamma R T}$ is the speed of sound. The initial Reynolds number based on the Taylor microscale $\lambda=u' / \langle ( \frac{\partial {u_i}}{\partial x_i} ) ^2 \rangle^{1/2} $ is $R_\lambda=\frac{\left\langle{\rho}\right\rangle u'\lambda}{\sqrt{3}\left\langle{\mu}\right\rangle}$=450.
Here, we use the initial large-eddy-turnover time $\tau_0 = (\int_0^\infty E(k)/k d k )/ u'$ to normalize time when quantifying the flow evolution.
The computational domain is discretized with a $512^3$ uniform grid, leading to a resolution down to the dissipation scale over the entire simulation duration: the ratio of the Kolmogorov scale $\eta$ over the grid size $\Delta$ grows from $\eta / \Delta = 0.98$ at $t=0$ to $\eta / \Delta = 2.88$ at $t=20$.

\begin{figure}
\begin{center}
\subfigure[]{
  \includegraphics[width=0.47\textwidth]{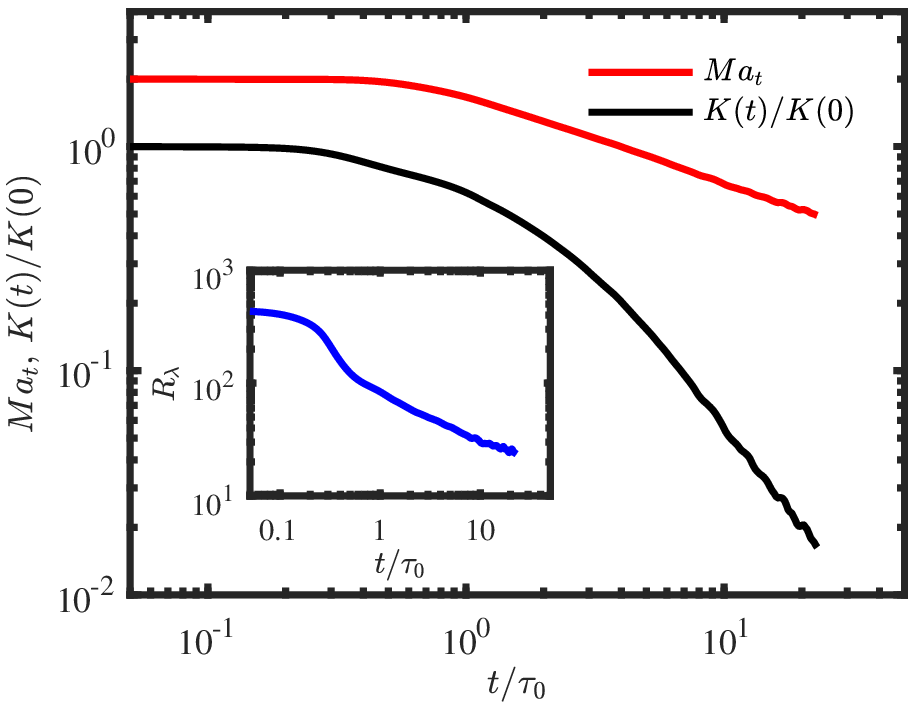}
}
\subfigure[]{
  \includegraphics[width=0.47\textwidth]{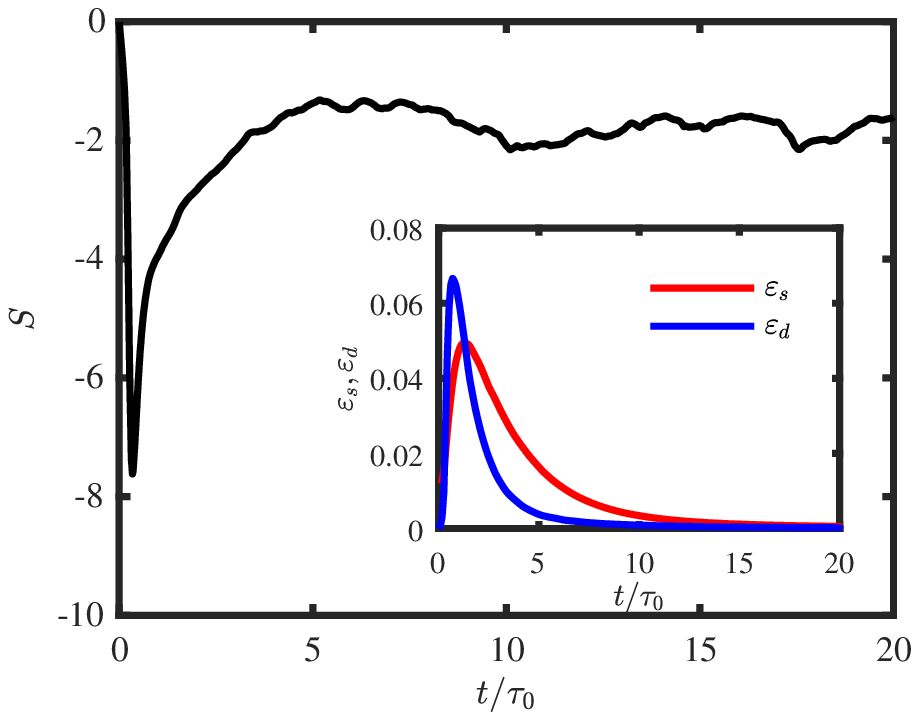}
}
\caption{ (a) Evolution of the turbulent Mach number $M_t$, the turbulence kinetic energy normalized with its initial value $K(t)/K(0)$, and the Reynolds number $R_\lambda$ (shown in inset). (b) The evolution of the skewness of the longitudinal velocity derivative $S$. The inset shows the energy dissipation rates: the solenoidal part $\varepsilon_s$ and the dilational part $\varepsilon_d$. 
}
\label{fig:tkeskew}
\end{center}
\end{figure}

Figure \ref{fig:tkeskew}a shows that the turbulent Mach number $M_t$, the
turbulent kinetic energy $K=\frac{1}{2}\langle \rho{u_i}{u_i}\rangle$, and the Reynolds number $R_\lambda$, all decay monotonically with time.
Note that even at time $t/\tau_0 > 10$, $Ma_t \lesssim 1$, the flow is still highly compressible with a large number of spatially distributed shocklets.
The skewness of the longitudinal velocity derivative, $S = \langle (  \partial {u_i}/\partial x_i )^3 \rangle / \langle (  \partial {u_i}/\partial x_i )^2 \rangle^{3/2}$ shown in Fig.~\ref{fig:tkeskew}b, however,
develops a sharp negative peak at $t/\tau_0 \approx 0.35$ and then gradually returns to a value fluctuating around $-2$, consistent with \cite{Samtaney417}. The larger magnitude of $S$ in the later stage compared with $S \approx -0.6$ in \cite{Samtaney417} is most likely due to the higher grid resolution in this work. In fact, \cite{Wang2012compressiso} demonstrated in DNS of forced compressible turbulence that for a given value of $R_\lambda$ and $M_t$, the magnitude of $S$ increases when the grid is refined. 
We note here that fully resolving the shocklets is not feasible as their sizes are out-of-reach of the current DNS with a fixed grid.
The peak of $S$ is a consequence of compressibility.
In the inset of Fig.~\ref{fig:tkeskew}b, we show the evolution of the turbulence dissipation rate, separated into the solenoidal (enstrophy) part $\varepsilon_s \equiv \langle {\mu \omega_{i}\omega_{i}}  \rangle $ and the dilational part $\varepsilon_d\equiv \frac{4}{3} \langle  \mu \overline{\m} ^2 \rangle$. 
The solenoidal dissipation $\varepsilon_s$ grows 
first due to the build-up of turbulent structures from the initial random field, reaches a peak at $t/\tau_0 \approx 0.8$,
then decreases gradually due to the decay of the turbulent fluctuation.
The dilational dissipation rate, $\varepsilon_d$,
represents the contribution of compressibility to dissipation,
which is exactly zero in incompressible turbulence.
With our choice of solenoidal initial condition,
$\varepsilon_d$ starts at zero and first grows rapidly, 
as shocklets are forming~\citep{Samtaney417}.
The peak of $\varepsilon_d$ is reached at $t/\tau_0 \approx 0.65$,  
slightly earlier than $\varepsilon_s$.
In our simulation, $\varepsilon_d$ contributes to approximately half the energy dissipation at the peak, but the solenoidal part remains the major cause of dissipation at later times.

\begin{figure}
\begin{center}

\subfigure[]{
  \includegraphics[width=0.47\textwidth]{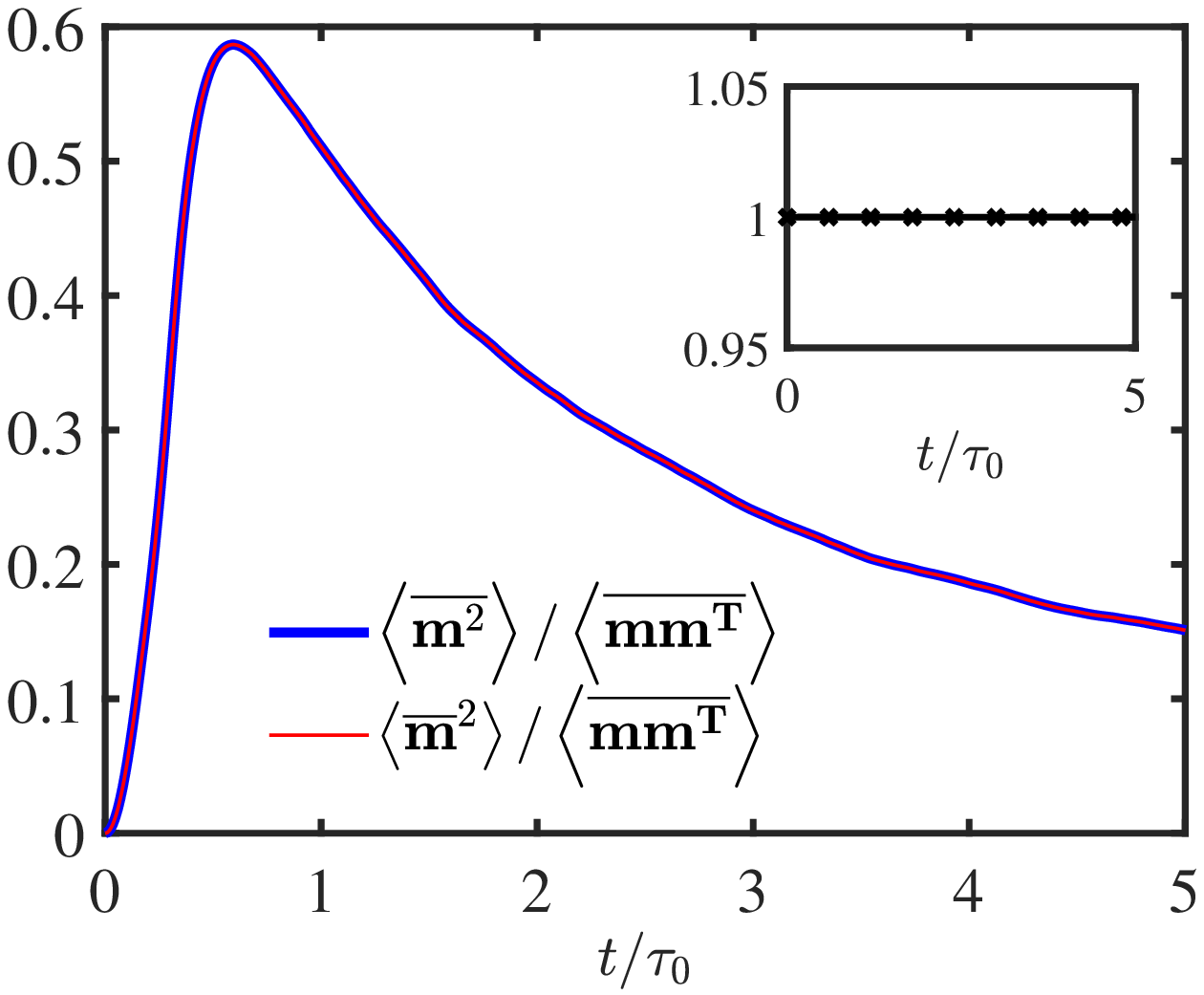}
}
\subfigure[]{
  \includegraphics[width=0.47\textwidth]{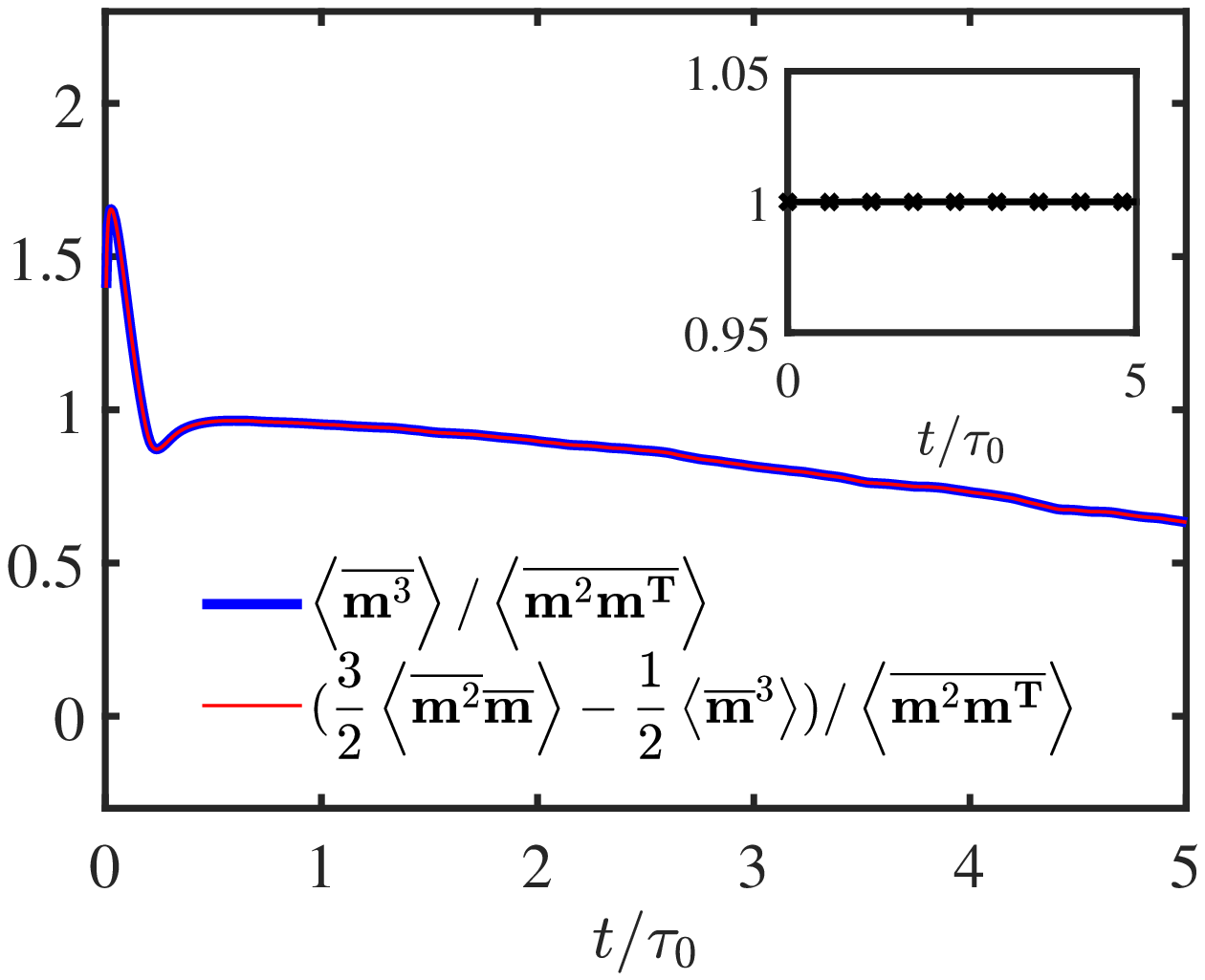}
}

\caption{Evolution of invariants of $\Aa^{(2)}$ and $\Aa^{(3)}$ in compressible homogeneous isotropic decaying turbulence. (a) Second-order invariants, $\langle \overline{\mathbf{m}^2} \rangle /\langle \overline{\mathbf{mm^T}} \rangle $ and $ \langle \overline {\mathbf{m}} ^2 \rangle / \langle \overline{\mathbf{mm^T}} \rangle $, the inset plots the ratio $ \langle \overline {\mathbf{m}} ^2 \rangle / \langle \overline{\mathbf{m}^2}\rangle $.
(b) Third-order invariants $\langle \overline {\mathbf{m}^3} \rangle /\langle \overline{\mathbf{m}^2 \mathbf{m^T}} \rangle  $ and $(\frac{3}{2}\langle \overline{\mathbf{m}^2} \overline{\mathbf{m}} \rangle - \frac{1}{2} \langle \overline {\mathbf{m}}^3 \rangle)  / \langle \overline{\mathbf{m}^2 \mathbf{m^T}} \rangle $, the inset plots the ratio $\langle \overline {\mathbf{m}^3}  \rangle /(\frac{3}{2} \langle \overline{\mathbf{m}^2} \overline{\mathbf{m}} \rangle - \frac{1}{2} \langle \overline {\mathbf{m}}^3 \rangle) $. 
}
\label{fig:betchov_compressible_HIT}
\end{center}
\end{figure}

We now discuss the structure of the velocity gradient correlations in this flow.
Figure~\ref{fig:betchov_compressible_HIT} shows the evolution of the l.h.s. and r.h.s. of Eq.~\eqref{eq:Betchov2compressible} and~\eqref{eq:Betchov3compressible}, the identities for the invariants of $\Aa^{(2)}$ and $\Aa^{(3)}$, respectively. The ratios between the two sides of those equations, shown in the insets, are exactly 1, as predicted for homogeneous flows.

\begin{figure}
\begin{center}
\subfigure[]{
  \includegraphics[width=0.47\textwidth]{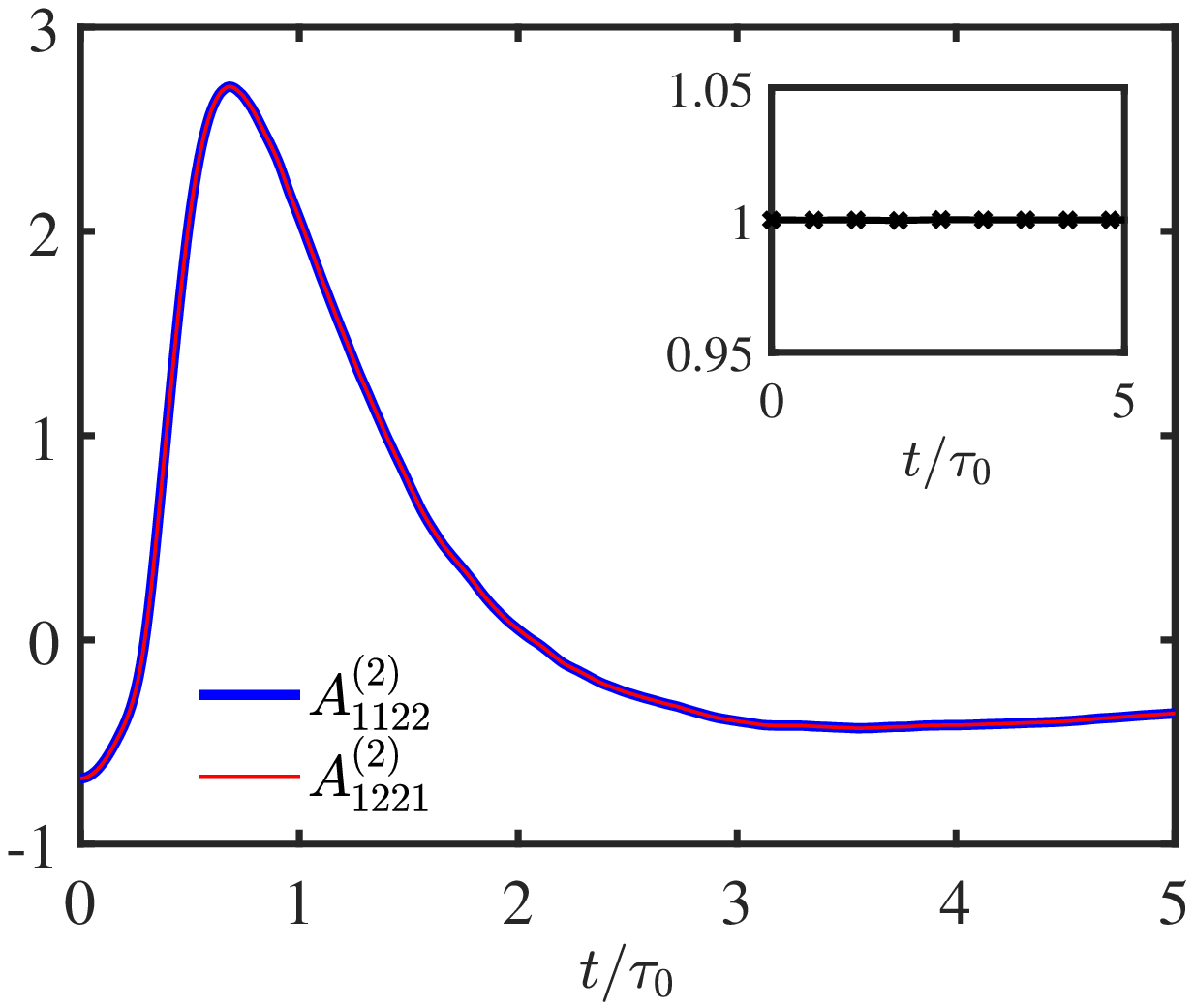}
}
\subfigure[]{
  \includegraphics[width=0.47\textwidth]{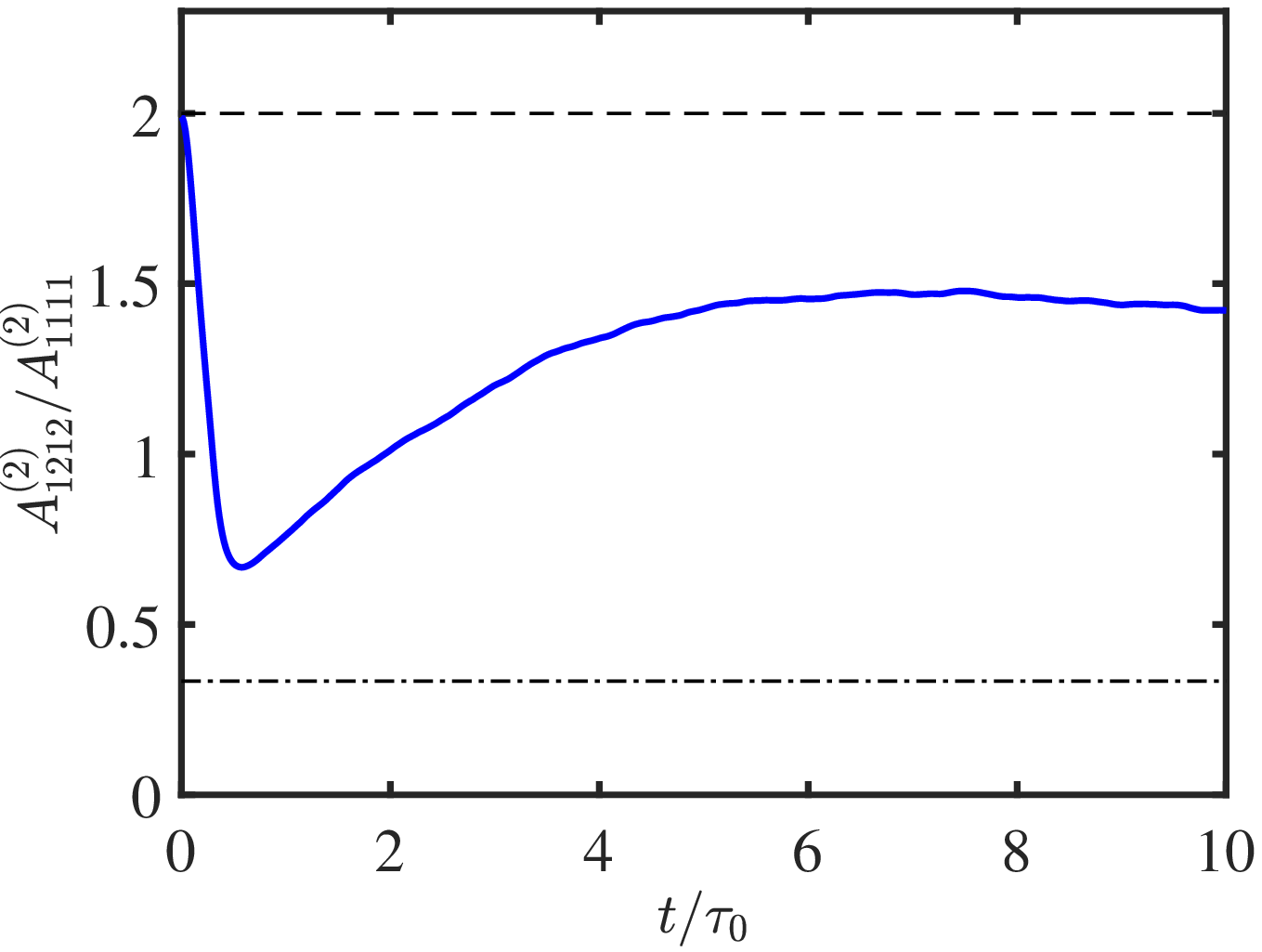}
}
\caption{ (a) Time evolution of $A^{(2)}_{1122}$ and $A^{(2)}_{1221}$ and their ratio (shown in inset). 
}
\label{fig:components_compressible_HIT}
\end{center}
\end{figure}

Our analysis predicts, for homogeneous and isotropic turbulence, further relations among the components of $\Aa^{(2)}$ and $\Aa^{(3)}$. 
Figure \ref{fig:components_compressible_HIT}a shows that $A^{(2)}_{1122}$ and $A^{(2)}_{1221}$ are equal during the entire simulation, as predicted by
Eq.~\eqref{eq:Aijkl_component_identity}.
In Fig.~\ref{fig:components_compressible_HIT}b, the ratio of the components $A^{(2)}_{1212} / A^{(2)}_{1111}$ also lies within $1/3$ and $2$ as predicted. It starts at $2$ since the flow is initially solenoidal. The rapid drop
to values lower than $1$ is concurrent with the rise of dilational dissipation $\varepsilon_d$. 
The minimal possible value of $1/3$ for $A^{(2)}_{1212} / A^{(2)}_{1111}$ corresponds to a compressible irrotational flow. The observed value of 
$A^{(2)}_{1212} / A^{(2)}_{1111}$, close to $1.5$ at later times, indicates 
instead the prevalence of the solenoidal part.

\begin{figure}
\begin{center}
\subfigure[]{
  \includegraphics[width=0.47\textwidth]{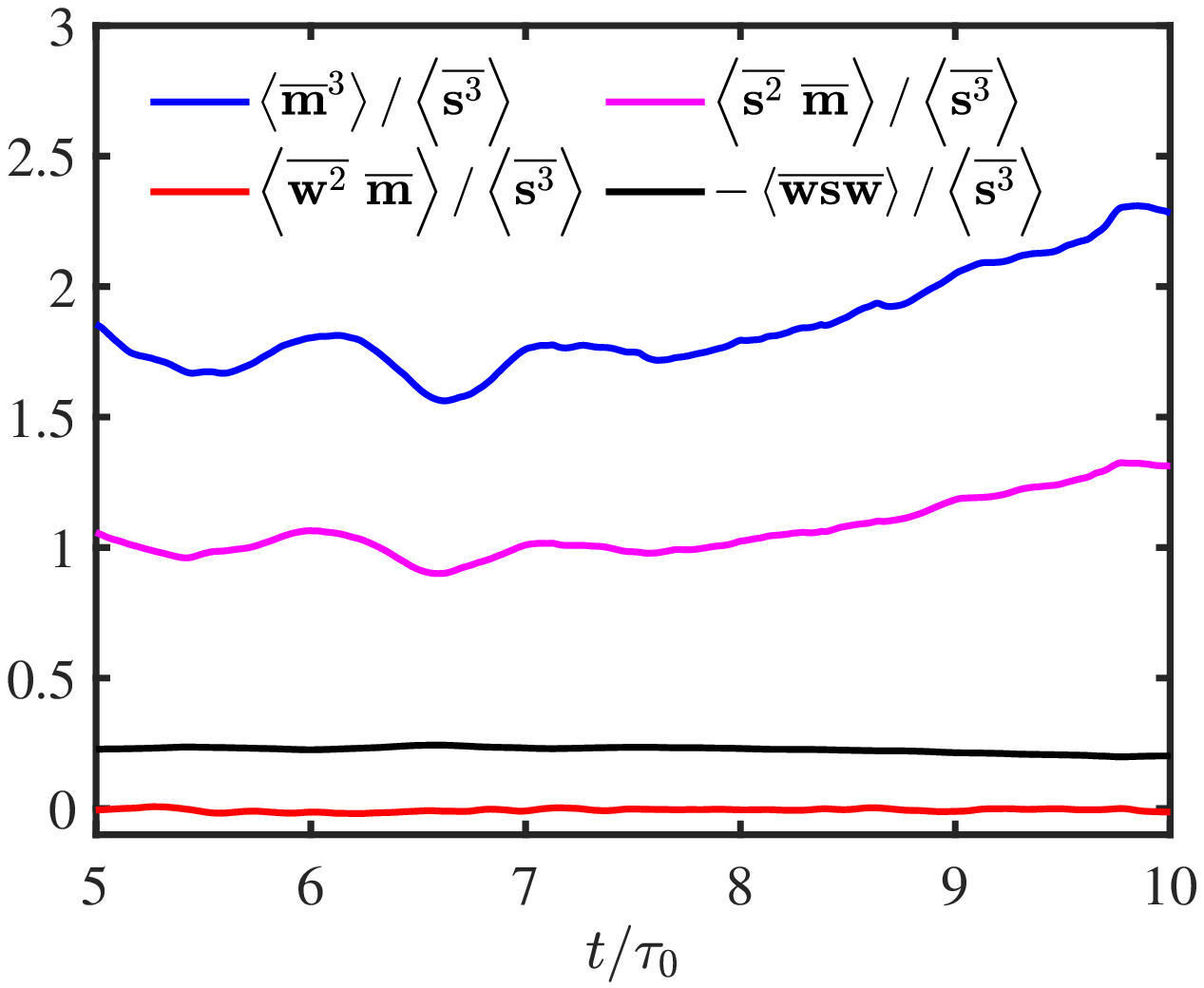}
}
\subfigure[]{
  \includegraphics[width=0.47\textwidth]{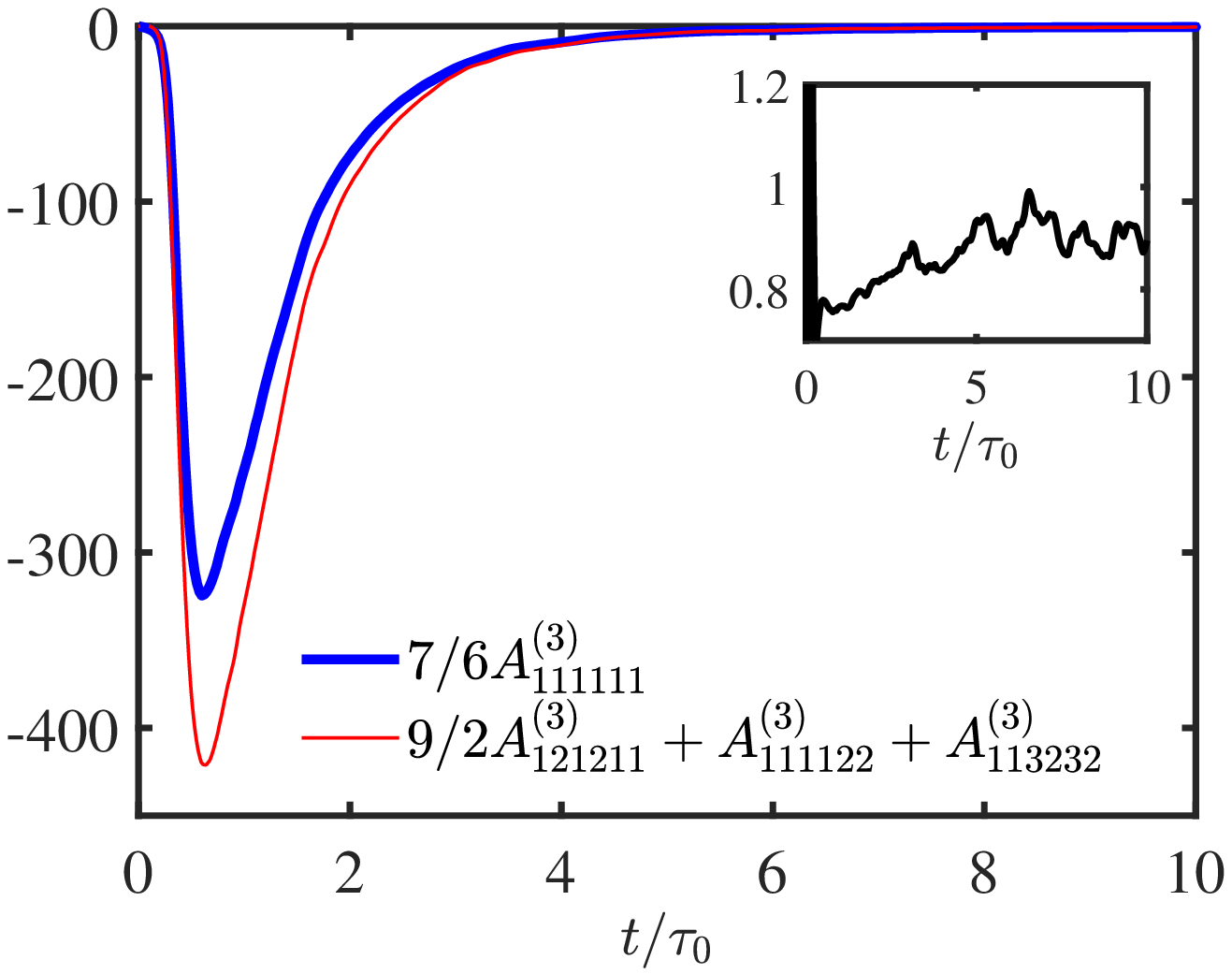}
}
\caption{ (a) Evolution of the invariants  $\langle \overline {\mathbf{m}^3}  \rangle$, $\langle \overline {\mathbf{w}^2}~\overline {\mathbf{m}} \rangle$, $\langle \overline {\mathbf{s}^2}~\overline {\mathbf{m}} \rangle$, and $\langle \overline{\mathbf{wsw}} \rangle$, all normalized by $\langle \overline {\mathbf{s}^3}  \rangle$. 
(b) The values of the l.h.s. and r.h.s. of Eq.~\eqref{eq:w2m_approx_1} and their ratio (shown in inset).
}
\label{fig:3rdinvariants_compressible_HIT}
\end{center}
\end{figure}

Figure \ref{fig:3rdinvariants_compressible_HIT}a shows the evolution of the invariants of $\Aa^{(3)}$ after the early stage when the flow rapidly adjusts in response to the initial divergence-free condition. Remarkably,
the magnitude of $\langle \overline {\mathbf{w}^2} \overline {\mathbf{m}} \rangle$ is very small compared with the others, consistent with the observation that in compressible homogeneous turbulence, vorticity and dilation are nearly uncorrelated \citep{Erlebacher1993CompressibleHT, Wang2012compressiso}.
This, in view of Eq.~\eqref{eq:A_iso_invariants_9},
provides an additional relation: $3 a_2 \approx 3 a_3 + 2 a_4 - 2 a_5$,
which then leaves only 3 independent quantities for a complete determination of $\Aa^{(3)}$.
Plugging Eq.~\eqref{eq:A_iso_as_1} to \eqref{eq:A_iso_as_4} into 
this relation leads to
\begin{equation}
\frac{7}{6} A_{111111} \approx \frac{9}{2} A_{121211} + A_{111122} + A_{113232}.
 \label{eq:w2m_approx_1}
\end{equation}
Figure \ref{fig:3rdinvariants_compressible_HIT}b
This result may be helpful to reconstruct the complete isotropic expression of $A^{(3)}_{ipjqkr}$ from planar PIV or 2C LDV data, instead of requiring stereo-PIV or 3C LDV.

Next we perform a DNS of a planar compressible mixing layer, formed by two co-moving free streams with Mach numbers $Ma_1=U_1/c_1=7.5$ and $Ma_2=U_2/c_2=1.5$, respectively, in which $U_1$ and $U_2$ are the mean velocities in the upper and lower free streams, and $c_1$ and $c_2$ are the speeds of sound in the two streams.
Similar with the setup of ~\cite{LiJaberi511}, at the inflow plane $x=0$, the mean stream velocity profile $U(0, y, z)$ is specified to be 
$U(0,y,z) =\frac{1}{2}\left[U_1+U_2 +(U_1-U_2)\tanh\left(2y/\delta_{\omega 0} \right)\right],$
with the inflow vorticity thickness ${\delta_{\omega 0}} = 1$. 
The convective Mach number is $Ma_c = (U_1-U_2)/(c_1+c_2)=3$ and the Reynolds number is $Re_c =\rho_1 (U_1-U_2){\delta_{\omega 0}}/\mu_1=3500$.
The mean temperature profile at the inlet is given by the Crocco-Busemann law with a uniform mean pressure. 
The effective size of the computational domain is $450{\delta_{\omega 0}}\times100{\delta_{\omega 0}}\times16{\delta_{\omega 0}}$  in the $x$, $y$ and $z$ directions, respectively. The domain is discretized using a mesh of $3050\times400\times128$ nodes
uniformly distributed in the  $x$ and $z$ directions and stretched in the $y$ direction with higher resolution in the centre of the mixing layer. 
Near the outflow plane, a sponge layer with a highly stretched mesh is added to damp fluctuations near the boundary. Random velocity fluctuations are superposed on the mean profile at the inlet plane to trigger turbulence, which develops downstream, forming 
a large number of shocklets in both upper and lower parts of the mixing layer.
When it is not too close to the inlet plane, the momentum thickness, $\theta$, grows linearly downstream  and the mean velocity profiles are self-similar, i.e., $U(x,y)$ at different $x$ locations collapse to $U(\xi)$ with $\xi=(y-y_c)/\theta$, where $y_c$ is the centre of the mixing layer.
The result is validated by checking the balance of turbulence kinetic energy budget (not shown). 

\begin{figure}
\begin{center}
\subfigure[]{
  \includegraphics[width=0.462\textwidth]{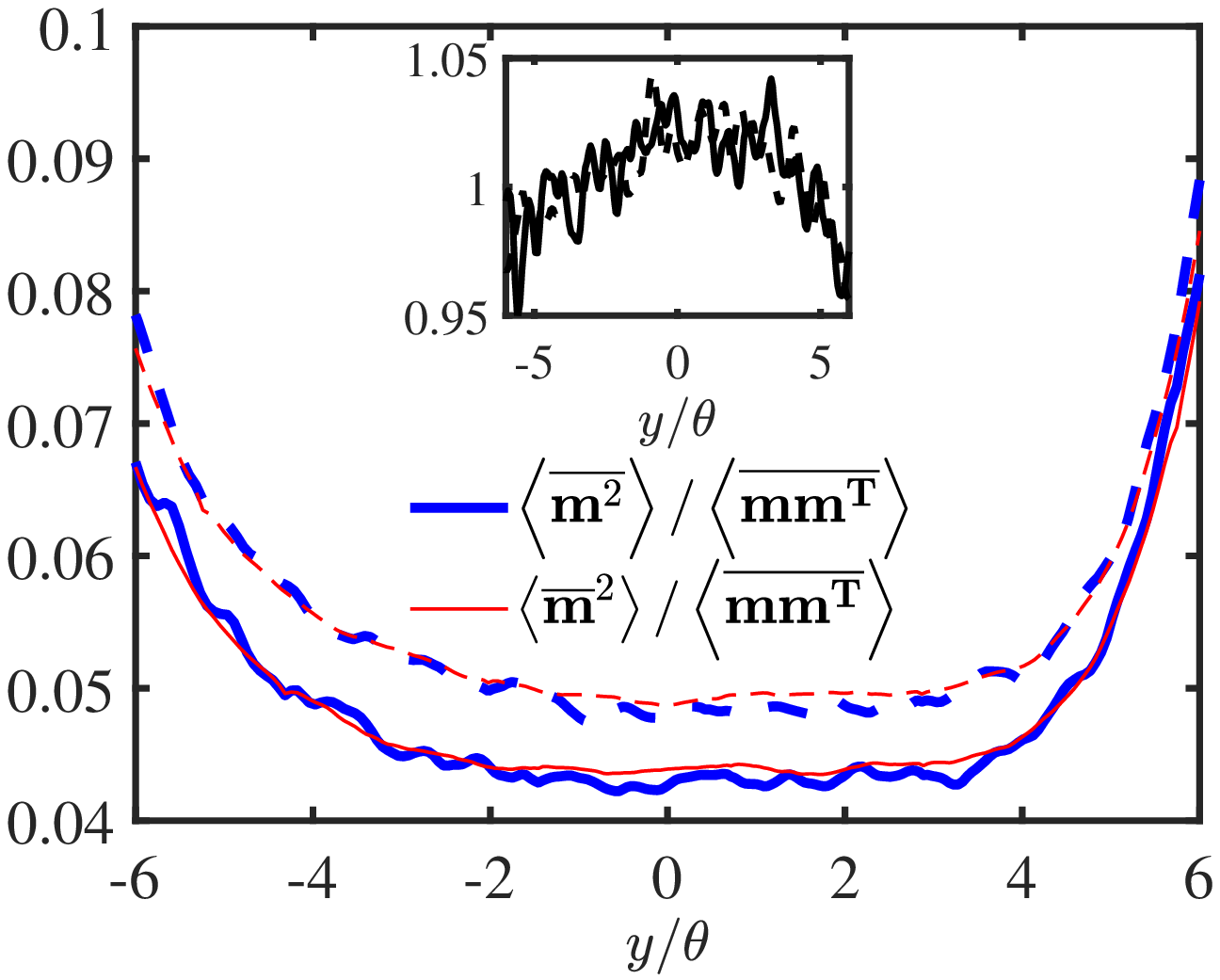}
}
\subfigure[]{
  \includegraphics[width=0.462\textwidth]{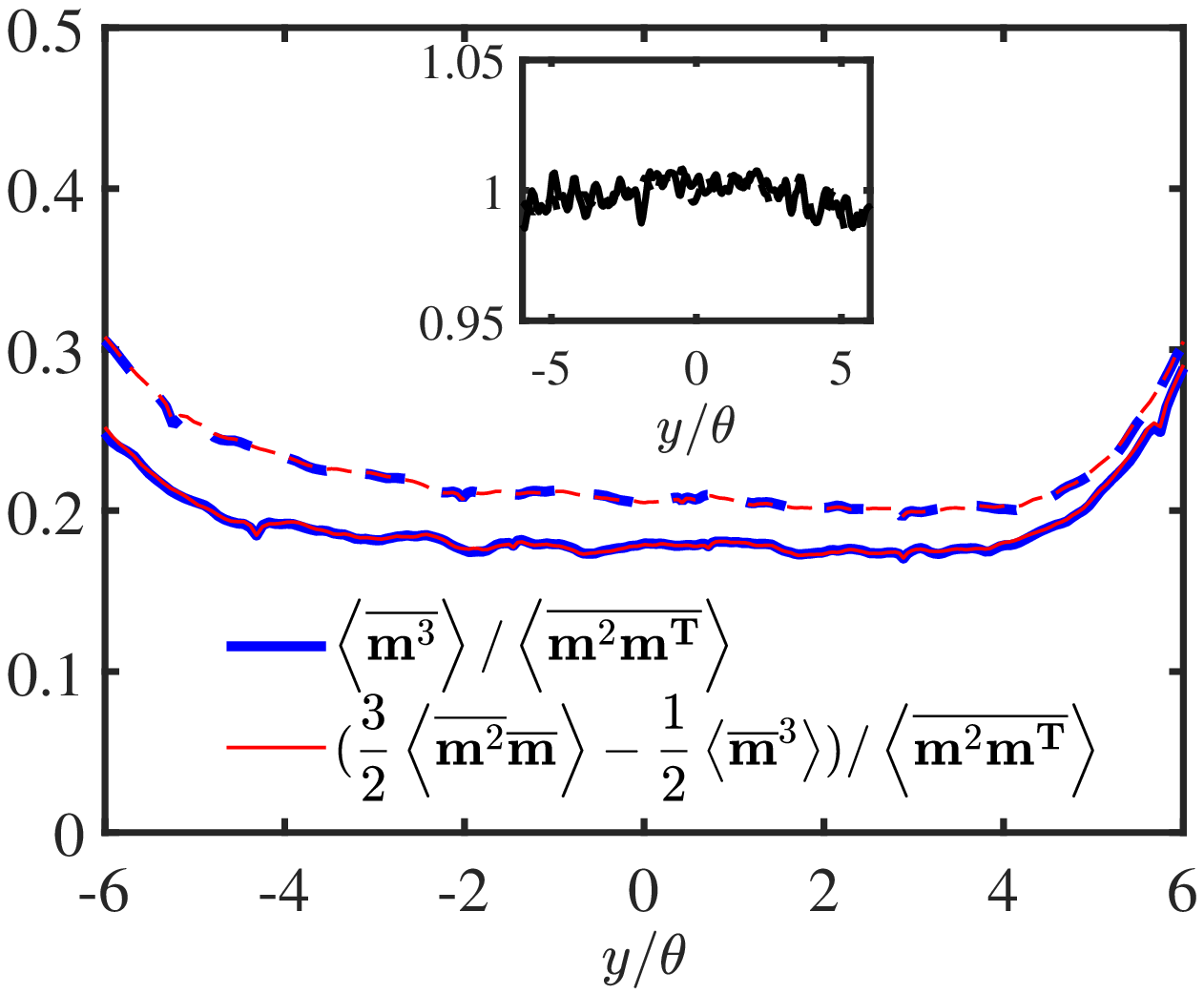}
}
\caption{Approximate validity of the homogeneous relations in the compressible mixing layer. (a) Normalized second-order invariants, $\langle \overline{\mathbf{m}^2} \rangle / \langle \overline{\mathbf{mm^T}}  \rangle $, $\langle \overline {\mathbf{m}}^2 \rangle / \langle \overline{\mathbf{mm^T}}  \rangle $, and their ratio (shown in the inset). (b) Normalized third-order invariants, $\langle \overline {\mathbf{m}^3}  \rangle / \langle \overline{\mathbf{m}^2 \mathbf{m^T}} \rangle $, $(\frac{3}{2}\langle \overline{\mathbf{m}^2} \overline{\mathbf{m}}  \rangle - \frac{1}{2} \langle \overline {\mathbf{m}}^3 \rangle)  / \langle \overline{\mathbf{m}^2 \mathbf{m^T}} \rangle $, and their ratio (shown in the inset). 
In both plots, dashed lines correspond to $x/\delta_{\omega 0} = 350$ and solid lines to $x/\delta_{\omega 0} = 400$.
}
\label{fig:betchov_compressible_ML}
\end{center}
\end{figure}

\begin{figure}
\begin{center}
\subfigure[]{
  \includegraphics[width=0.462\textwidth]{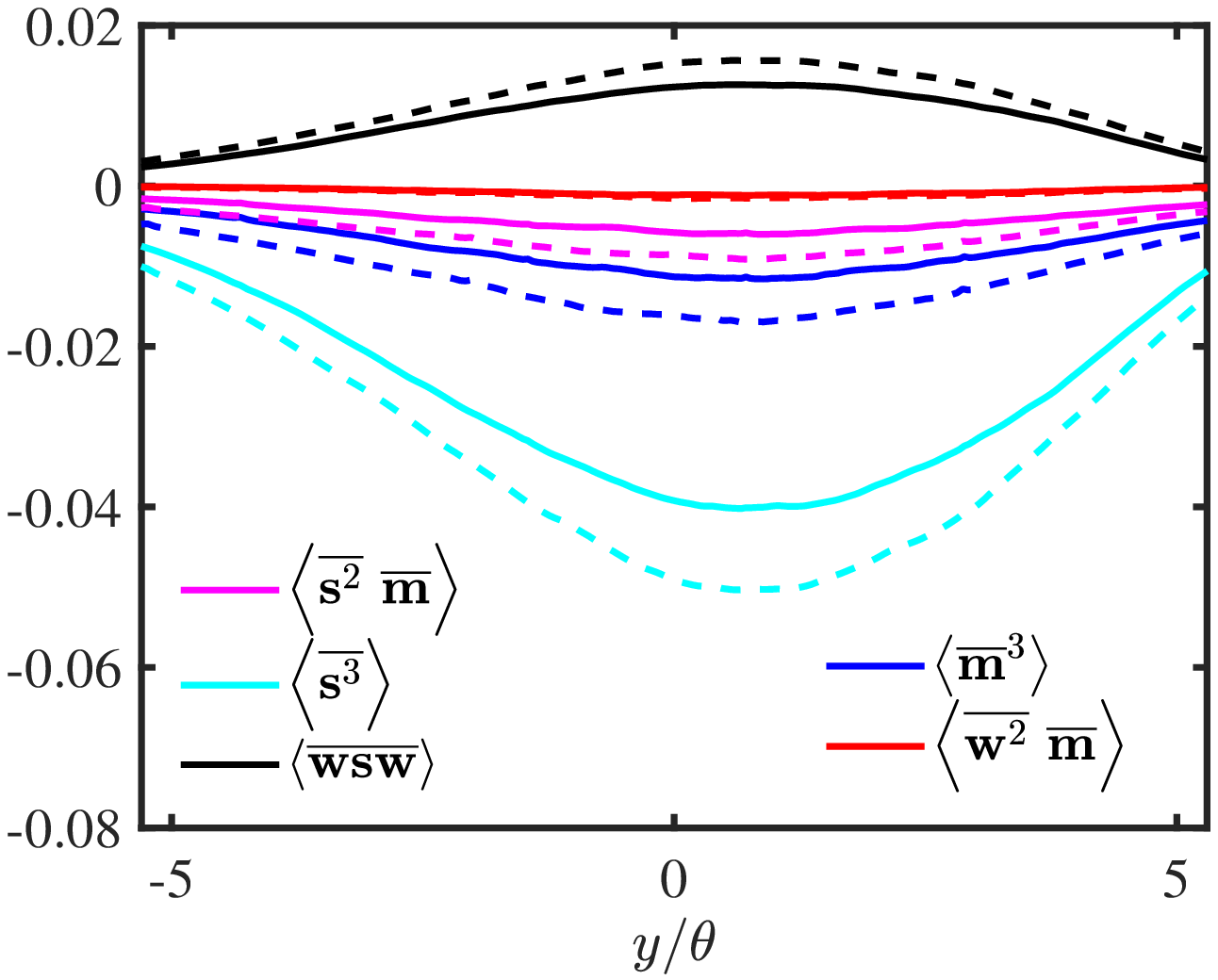}
}
\subfigure[]{
  \includegraphics[width=0.462\textwidth]{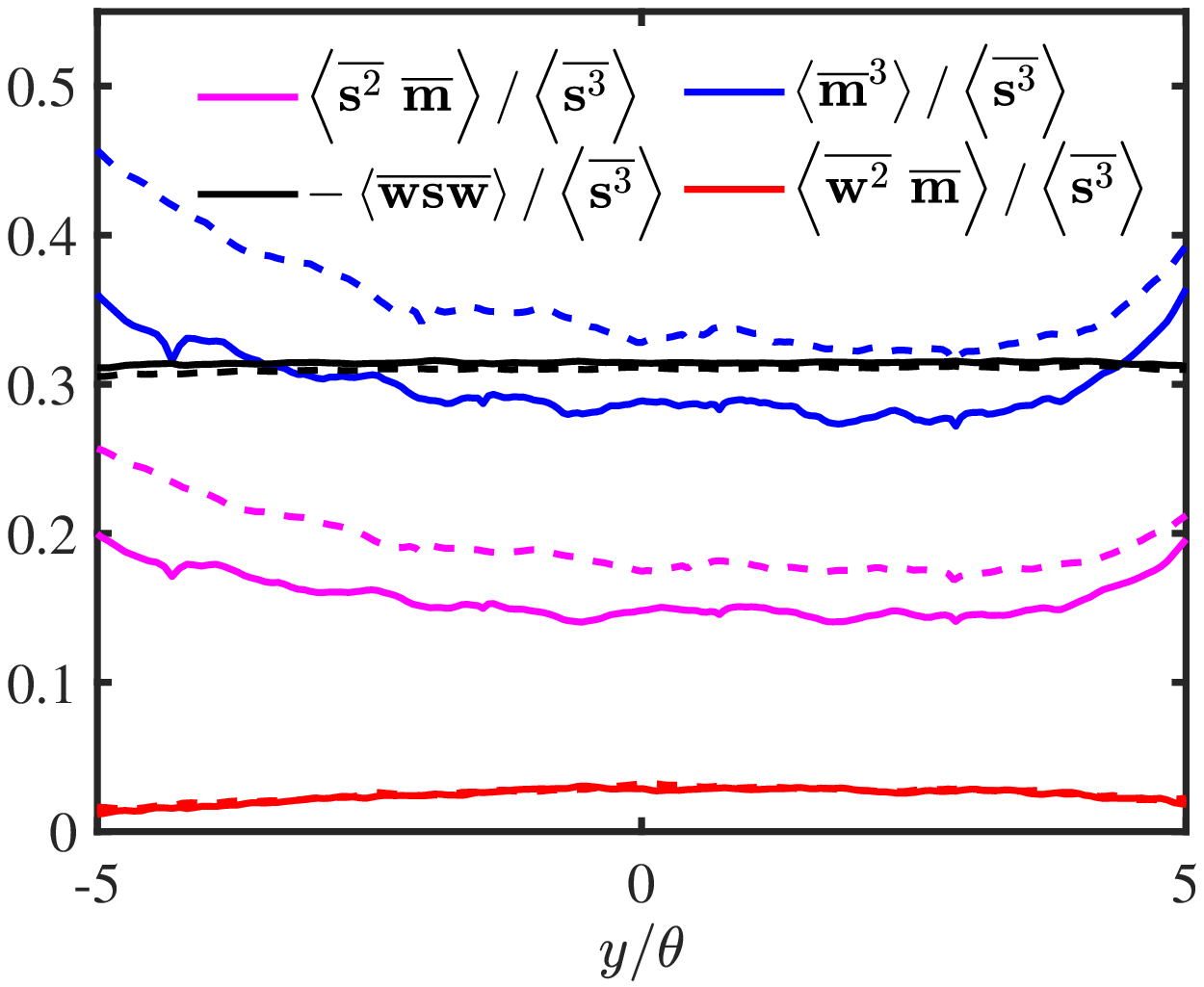}
}
\caption{ (a) Profiles of the invariants $\langle \overline{\m}^3 \rangle$, $\langle \overline{\mathbf{w}^2}\overline {\m} \rangle$, $\langle \overline{\mathbf{s}^2} \overline{\mathbf{m}} \rangle$, $\langle \overline{\mathbf{s}^3}  \rangle$ and $\langle \overline{\mathbf{wsw}} \rangle$ in the mixing layer.  (b) Relative ratios  $\langle \overline{\mathbf{m}^3} \rangle$/$\langle \overline{\mathbf{s}^3} \rangle$, $\langle \overline{\mathbf{w}^2} \overline{\mathbf{m}} \rangle$/$\langle \overline{\mathbf{s}^3} \rangle$, $\langle \overline{\mathbf{s}^2} \overline{\mathbf{m}} \rangle$/$\langle \overline{\mathbf{s}^3} \rangle$, and $\langle \overline{\mathbf{wsw}} \rangle$/$\langle \overline{\mathbf{s}^3} \rangle$.
In both plots, dashed lines correspond to $x/\delta_{\omega 0} = 350$ and solid lines to $x/\delta_{\omega 0} = 400$.
}
\label{fig:3rdinvariants_compressible_channel}
\end{center}
\end{figure}

Figure \ref{fig:betchov_compressible_ML} shows the ratio between the l.h.s. and the r.h.s. of Eq.~\eqref{eq:Betchov2compressible} and \eqref{eq:Betchov3compressible} in the center region of the mixing layer $(-6 \leqslant y/\theta \leqslant 6)$ at the downstream locations $x\delta_{\omega 0} = 350$ (dashed lines) and $x / \delta_{\omega 0}=400$ (solid lines). 
Although the flow is not homogeneous, the ratios $\langle \overline {\m} ^2 \rangle / \langle \overline{\mathbf{m}^2} \rangle $ and $ \langle \overline {\mathbf{m}^3}  \rangle /(\frac{3}{2} \langle \overline{\mathbf{m}^2} \overline{\m} \rangle - \frac{1}{2} \langle \overline {\mathbf{m}}^3 \rangle)$ are very close to unity, so Eqs.~\eqref{eq:Betchov2compressible} and \eqref{eq:Betchov3compressible} are still approximately valid.
Fig.~\ref{fig:3rdinvariants_compressible_channel} shows the profiles of various third-order invariants $\langle \overline{\m}^3 \rangle$, $\langle \overline{\mathbf{w}^2}\overline {\m} \rangle$, $\langle \overline{\mathbf{s}^2} \overline{\mathbf{m}} \rangle$, $\langle \overline{\mathbf{s}^3}  \rangle$ and $\langle \overline{\mathbf{wsw}} \rangle$. Despite the inhomogeneity and anisotropy, the vorticity-dilation correlation $\langle \overline{\mathbf{w}^2}\overline {\m} \rangle$ remains very small compared to other invariants, which could help to obtain more 
relations among invariants.

\section{Concluding Remarks}
In summary, we derived exact relations among invariants of the moments of velocity gradients in compressible homogeneous turbulence and verified these relations by DNS of decaying compressible turbulence. Interestingly, these relations, derived under homogeneity assumptions, hold approximately in a compressible mixing layer. We also devised approaches 
to determine the full tensor with a minimal set of measurements in experiments. 
These relations could help, e.g., to determine separately the solenoidal and the dilational energy dissipation rates from only two velocity derivatives $\frac{\partial u_1}{\partial x_1}$ and $\frac{\partial u_2}{\partial x_1}$.
In the future, it would be interesting to investigate the structure of the velocity gradient implied by these relations, as done by \cite{Betchov-1956} for incompressible turbulence.

\section*{Acknowledgements}
PFY and HX acknowledge support from the
National Natural Science Foundation of China (NSFC) under grants 11672157 and 91852104.
JF  acknowledges the UK Engineering and Physical
Sciences Research Council (EPSRC) through the Computational
Science Centre for Research Communities (CoSeC), and the UK
Turbulence Consortium (No. EP/R029326/1).
AP was supported by the French Agence National de la Recherche under Contract No. ANR-20-CE30-0035 (project TILT).
The simulations were conducted on the ARCHER2 UK National Supercomputing Service.

\noindent
\textbf{Declaration of Interest:} 
The authors report no conflict of interest.

\section*{Appendix}
\label{sec:Appendix}

For vector gradients $\frac{\partial a_i}{\partial x_j}$, $\frac{\partial b_i}{\partial x_j}$ and $\frac{\partial c_i}{\partial x_j}$, we have, by elementary algebra:
\begin{align}
\left \langle \frac{\partial a_i}{\partial x_j} \frac{\partial b_j}{\partial x_k} \frac{\partial c_k}{\partial x_i} \right\rangle &= \frac{\partial}{\partial x_j} \left\langle a_i \frac{\partial b_j}{\partial x_k} \frac{\partial c_k}{\partial x_i}  \right\rangle - \left \langle a_i \frac{\partial^2 b_j}{\partial x_j \partial x_k} \frac{\partial c_k}{\partial x_i} \right\rangle - \left \langle a_i \frac{\partial b_j}{\partial x_k} \frac{\partial^2 c_k}{\partial x_i \partial x_j} \right\rangle 
\label{eq:Ap1}\\
\left \langle a_i \frac{\partial^2 b_j}{\partial x_j \partial x_k} \frac{\partial c_k}{\partial x_i} \right\rangle & = \frac{\partial}{\partial x_k} \left\langle a_i \frac{\partial b_j}{\partial x_j} \frac{\partial c_k}{\partial x_i}  \right\rangle - \left \langle \frac{\partial a_i}{\partial x_k} \frac{\partial b_j}{\partial x_j} \frac{\partial c_k}{\partial x_i} \right\rangle - \left \langle a_i \frac{\partial b_j}{\partial x_j} \frac{\partial^2 c_k}{\partial x_k \partial x_i} \right\rangle  
\label{eq:Ap2}\\
\left \langle a_i \frac{\partial b_j}{\partial x_k} \frac{\partial^2 c_k}{\partial x_i \partial x_j} \right\rangle &= \frac{\partial}{\partial x_i} \left\langle a_i \frac{\partial b_j}{\partial x_k} \frac{\partial c_k}{\partial x_j}  \right\rangle - \left \langle \frac{\partial a_i}{\partial x_i} \frac{\partial b_j}{\partial x_k} \frac{\partial c_k}{\partial x_j} \right\rangle - \left \langle a_i \frac{\partial^2 b_j}{\partial x_i \partial x_k} \frac{\partial c_k}{\partial x_j} \right\rangle \label{eq:Ap3} \\
\left \langle a_i \frac{\partial^2 b_j}{\partial x_i \partial x_k} \frac{\partial c_k}{\partial x_j} \right\rangle &= \frac{\partial}{\partial x_k} \left\langle a_i \frac{\partial b_j}{\partial x_i} \frac{\partial c_k}{\partial x_j}  \right\rangle - \left \langle \frac{\partial a_i}{\partial x_k} \frac{\partial b_j}{\partial x_i} \frac{\partial c_k}{\partial x_j} \right\rangle - \left \langle a_i \frac{\partial b_j}{\partial x_i} \frac{\partial^2 c_k}{\partial x_k \partial x_j} \right\rangle  \label{eq:Ap4}\\
\left \langle a_i \frac{\partial b_j}{\partial x_j} \frac{\partial^2 c_k}{\partial x_k \partial x_i} \right\rangle &= \frac{\partial}{\partial x_i} \left\langle a_i \frac{\partial b_j}{\partial x_j} \frac{\partial c_k}{\partial x_k}  \right\rangle - \left \langle \frac{\partial a_i}{\partial x_i} \frac{\partial b_j}{\partial x_j} \frac{\partial c_k}{\partial x_k} \right\rangle - \left \langle a_i \frac{\partial^2 b_j}{\partial x_i \partial x_j} \frac{\partial c_k}{\partial x_k} \right\rangle \label{eq:Ap5} \\
\left \langle a_i \frac{\partial b_j}{\partial x_i} \frac{\partial^2 c_k}{\partial x_k \partial x_j} \right\rangle &= \frac{\partial}{\partial x_j} \left\langle a_i \frac{\partial b_j}{\partial x_i} \frac{\partial c_k}{\partial x_k}  \right\rangle - \left \langle \frac{\partial a_i}{\partial x_j} \frac{\partial b_j}{\partial x_i} \frac{\partial c_k}{\partial x_k} \right\rangle - \left \langle a_i \frac{\partial^2 b_j}{\partial x_i \partial x_j} \frac{\partial c_k}{\partial x_k} \right\rangle \, . \label{eq:Ap6}
\end{align}
Multiplying (\ref{eq:Ap2}), (\ref{eq:Ap3}) and (\ref{eq:Ap6}) by $-1$, and summing over, one obtains:
\begin{equation}
2 \langle \overline{\mathbf{h^a} \mathbf{h^b} \mathbf{h^c}} \rangle  =  \langle \overline{\mathbf{h^a} \mathbf{h^b}} \, \overline{\mathbf{h^c}} \rangle + \langle \overline{\mathbf{h^b} \mathbf{h^c}} \, \overline{\mathbf{h^a}} \rangle + \langle \overline{\mathbf{h^c} \mathbf{h^a}} \, \overline{\mathbf{h^b}} \rangle - \langle \overline{\mathbf{h^a}} \, \overline{\mathbf{h^b}}  \, \overline{\mathbf{h^c}} \rangle ,
\label{eq:third_order_mixrelations}
\end{equation}
in which $\mathbf{h^a} = \nabla \mathbf{a}$ etc. For divergence-free fields, this yields $\langle \overline{\nabla \mathbf{a} \nabla \mathbf{b} \nabla \mathbf{c} } \rangle = 0$. An equivalent form of this special case has been shown in Appendix D of \cite{Eyink2006}.

\bibliographystyle{jfm}
%\bibliography{BIB}

\begin{thebibliography}{35}
\expandafter\ifx\csname natexlab\endcsname\relax\def\natexlab#1{#1}\fi
\def\au#1{#1} \def\ed#1{#1} \def\yr#1{#1}\def\at#1{#1}\def\jt#1{\textit{#1}}
  \def\bt#1{#1}\def\bvol#1{\textbf{#1}} \def\vol#1{#1} \def\pg#1{#1}
  \def\publ#1{#1}\def\arxiv#1{#1}\def\org#1{#1}\def\st#1{\textit{#1}}

\bibitem[Betchov(1956)]{Betchov-1956}
{\sc \au{Betchov, R.}} \yr{1956}  \at{An inequality concerning the production
  of vorticity in isotropic turbulence}.  \jt{J. Fluid Mech.}  \bvol{1},
  \pg{497--504}.

\bibitem[Bradshaw \& Perot(1993)]{Bradshaw1993}
{\sc \au{Bradshaw, P.} \& \au{Perot, J.~B.}} \yr{1993}  \at{A note on turbulent
  energy dissipation in the viscous wall region}.  \jt{Phys. Fluids A}
  \bvol{5},  \pg{3305--3306}.

\bibitem[Buaria {\em et~al.\/}(2020)Buaria, Pumir \& Bodenschatz]{Buaria20a}
{\sc \au{Buaria, D.}, \au{Pumir, A.} \& \au{Bodenschatz, E.}} \yr{2020}
  \at{{Vortex stretching and enstrophy production in high Reynolds number
  turbulence}}.  \jt{Phys. Rev. Fluids}  \bvol{5},  \pg{104602}.

\bibitem[Chen {\em et~al.\/}(2019)Chen, Wang, Li, Wan \&
  Chen]{ChenWang2019compressshear}
{\sc \au{Chen, S.}, \au{Wang, J.}, \au{Li, H.}, \au{Wan, M.} \& \au{Chen, S.}}
  \yr{2019}  \at{Effect of compressibility on small scale statistics in
  homogeneous shear turbulence}.  \jt{Phys. Fluids}  \bvol{31},  \pg{025107}.

\bibitem[Chu {\em et~al.\/}(2014)Chu, Wang \& Lu]{Chu2014compressibleBL}
{\sc \au{Chu, Y.~B.}, \au{Wang, L.} \& \au{Lu, X.~Y.}} \yr{2014}
  \at{Interaction between strain and vorticity in compressible turbulent
  boundary layer}.  \jt{Sci. China Phys., Mech. Astron.}  \bvol{57},
  \pg{2316--2329}.

\bibitem[Erlebacher \& Sarkar(1993)]{Erlebacher1993CompressibleHT}
{\sc \au{Erlebacher, G.} \& \au{Sarkar, S.}} \yr{1993}  \at{Rate of strain
  tensor statistics in compressible homogeneous turbulence}.  \jt{Phys. Fluids
  A}  \bvol{5},  \pg{3240--3254}.

\bibitem[Eyink(2006)]{Eyink2006}
{\sc \au{Eyink, G.}} \yr{2006}  \at{Multi-scale gradient expansion of the
  turbulent stress tensor}.  \jt{J. Fluid Mech.}  \bvol{549},  \pg{159--190}.

\bibitem[Fang {\em et~al.\/}(2019)Fang, Gao, Moulinec \& Emerson]{FangGao1850}
{\sc \au{Fang, J.}, \au{Gao, F.}, \au{Moulinec, C.} \& \au{Emerson, D.~R.}}
  \yr{2019}  \at{An improved parallel compact scheme for domain-decoupled
  simulation of turbulence}.  \jt{Int. J. for Numer. Meth. Fl.}
  \bvol{90}~(10),  \pg{479--500}.

\bibitem[Fang {\em et~al.\/}(2013)Fang, Li \& Lu]{FangLi886}
{\sc \au{Fang, J.}, \au{Li, Z.} \& \au{Lu, L.}} \yr{2013}  \at{An optimized
  low-dissipation monotonicity-preserving scheme for numerical simulations of
  high-speed turbulent flows}.  \jt{J. Sci. Comput.}  \bvol{56},  \pg{67--95}.

\bibitem[Fang {\em et~al.\/}(2014)Fang, Yao, Li \& Lu]{FangYao1856}
{\sc \au{Fang, J.}, \au{Yao, Y.}, \au{Li, Z.} \& \au{Lu, L.\textbf{}}}
  \yr{2014}  \at{Investigation of low-dissipation monotonicity-preserving
  scheme for direct numerical simulation of compressible turbulent flows}.
  \jt{Comput. Fluids}  \bvol{104},  \pg{55--72}.

\bibitem[Fang {\em et~al.\/}(2015)Fang, Yao, Zheltovodov, Li \& Lu]{FangYao977}
{\sc \au{Fang, J.}, \au{Yao, Y.}, \au{Zheltovodov, A.~A.}, \au{Li, Z.} \&
  \au{Lu, L.}} \yr{2015}  \at{Direct numerical simulation of supersonic
  turbulent flows around a tandem expansion-compression corner}.  \jt{Phys.
  Fluids}  \bvol{12}~(27).

\bibitem[Fang {\em et~al.\/}(2020)Fang, Zheltovodov, Yao, Moulinec \&
  Emerson]{Fang2009}
{\sc \au{Fang, J.}, \au{Zheltovodov, A.~A.}, \au{Yao, Y.}, \au{Moulinec, C.} \&
  \au{Emerson, D.~R.}} \yr{2020}  \at{On the turbulence amplification in
  shock-wave/turbulent boundary layer interaction}.  \jt{J. Fluid Mech.}
  \bvol{897},  \pg{A32}.

\bibitem[Fang {\em et~al.\/}(2016)Fang, Zhang, Fang \&
  Zhu]{Fang2016Fourthorder}
{\sc \au{Fang, L.}, \au{Zhang, Y.~J.}, \au{Fang, J.} \& \au{Zhu, Y.}} \yr{2016}
   \at{Relation of the fourth-order statistical invariants of velocity gradient
  tensor in isotropic turbulence}.  \jt{Phys. Rev. E}  \bvol{94},  \pg{023114}.

\bibitem[Frisch(1995)]{Frisch95}
{\sc \au{Frisch, U.}} \yr{1995} {\em Turbulence: The legacy of A. N.
  Kolmogorov\/}.  \publ{Cambridge Univ. Press.}

\bibitem[Gottlieb \& Shu(1998)]{GottliebShu522}
{\sc \au{Gottlieb, S.} \& \au{Shu, C.~W.}} \yr{1998}  \at{Total variation
  diminishing runge-kutta schemes}.  \jt{Math. Comput.}  \bvol{67},
  \pg{73--85}.

\bibitem[Lele(1992)]{Lele429}
{\sc \au{Lele, Sanjiva~K.}} \yr{1992}  \at{Compact finite difference schemes
  with spectral-like resolution}.  \jt{J. Comput. Phys.}  \bvol{103},
  \pg{16--42}.

\bibitem[Li \& Jaberi(2011)]{LiJaberi511}
{\sc \au{Li, Z.} \& \au{Jaberi, F.~A.}} \yr{2011}  \at{A high-order finite
  difference method for numerical simulations of supersonic turbulent flows}.
  \jt{J. Numer. Meth. Fluids}  \bvol{68}~(6),  \pg{740--766}.

\bibitem[Ma \& Xiao(2016)]{Ma2016compressibleshear}
{\sc \au{Ma, Z.} \& \au{Xiao, Z.}} \yr{2016}  \at{Turbulent kinetic energy
  production and flow structures in compressible homogeneous shear flow}.
  \jt{Phys. Fluids}  \bvol{28},  \pg{096102}.

\bibitem[Meneveau(2011)]{Meneveau2011ANNU}
{\sc \au{Meneveau, C.}} \yr{2011}  \at{Lagrangian dynamics and models of the
  velocity gradient tensor in turbulent flows}.  \jt{Annu. Rev. Fluid Mech.}
  \bvol{43},  \pg{219--245}.

\bibitem[Pan \& Johnsen(2017)]{Pan17}
{\sc \au{Pan, S.} \& \au{Johnsen, E.}} \yr{2017}  \at{{The role of bulk
  viscosity on the decay of compressible homogeneous, isotropic turbulence}}.
  \jt{J. Fluid Mech.}  \bvol{833},  \pg{717--744}.

\bibitem[Pirozzoli \& Grasso(2004)]{Pirozzoli2004isoCompressible}
{\sc \au{Pirozzoli, S.} \& \au{Grasso, F.}} \yr{2004}  \at{Direct numerical
  simulations of isotropic compressible turbulence: Influence of
  compressibility on dynamics and structures}.  \jt{Phys. Fluids}  \bvol{16},
  \pg{4386--4407}.

\bibitem[Pope(2000)]{Pope2000}
{\sc \au{Pope, S.~B.}} \yr{2000} {\em Turbulent flows\/}.  \publ{Cambridge
  University Press}.

\bibitem[Pumir(2017)]{pumir2017structure}
{\sc \au{Pumir, A.}} \yr{2017}  \at{Structure of the velocity gradient tensor
  in turbulent shear flows}.  \jt{Phys. Rev. Fluids}  \bvol{2}~(7),
  \pg{074602}.

\bibitem[Pumir {\em et~al.\/}(2016)Pumir, Xu \& Siggia]{PumirXu-2016}
{\sc \au{Pumir, A.}, \au{Xu, H.} \& \au{Siggia, E.~D.}} \yr{2016}
  \at{Small-scale anisotropy in turbulent boundary layers}.  \jt{J. Fluid
  Mech.}  \bvol{804},  \pg{5--23}.

\bibitem[Samtaney {\em et~al.\/}(2001)Samtaney, Pullin \& Kosovic]{Samtaney417}
{\sc \au{Samtaney, R.}, \au{Pullin, D.~I.} \& \au{Kosovic, B.}} \yr{2001}
  \at{Direct numerical simulation of decaying compressible turbulence and
  shocklet statistics}.  \jt{Phys. Fluids}  \bvol{13}~(5),  \pg{1415--1430}.

\bibitem[Sreenivasan \& Antonia(1997)]{Sreeni97}
{\sc \au{Sreenivasan, K.~R.} \& \au{Antonia, R.~A.}} \yr{1997}  \at{{The
  phenomenology of small-scale turbulence}}.  \jt{Annu. Rev. Fluid Mech.}
  \bvol{29},  \pg{435--472}.

\bibitem[Suman \& Girimaji(2009)]{Suman2009}
{\sc \au{Suman, S.} \& \au{Girimaji, S.~S.}} \yr{2009}  \at{{Homogenized Euler
  equation: a model for compressible velocity gradient dynamics}}.  \jt{J.
  Fluid Mech.}  \bvol{620},  \pg{177--194}.

\bibitem[Suman \& Girimaji(2011)]{Suman2011}
{\sc \au{Suman, S.} \& \au{Girimaji, S.~S.}} \yr{2011}  \at{{Dynamical model
  for velocity-gradient evolution in compressible turbulence}}.  \jt{J. Fluid
  Mech.}  \bvol{683},  \pg{289--319}.

\bibitem[Suman \& Girimaji(2013)]{Suman2013}
{\sc \au{Suman, S.} \& \au{Girimaji, S.~S.}} \yr{2013}  \at{{Velocity gradient
  dynamics in compressible turbulence: Characterization of pressure-Hessian
  tensor}}.  \jt{Phys. Fluids}  \bvol{25},  \pg{125103}.

\bibitem[Tsinober(2009)]{Tsinober2009}
{\sc \au{Tsinober, A.}} \yr{2009} {\em An Informal Conceptual Introduction to
  Turbulence\/}.  \publ{Springer}.

\bibitem[Vaghel \& Madnia(2015)]{Vaghel2015}
{\sc \au{Vaghel, N.S.} \& \au{Madnia, C.K.}} \yr{2015}  \at{Local flow topology
  and velocity gradient invariants in compressible turbulent mixing layer}.
  \jt{J. Fluid Mech.}  \bvol{774},  \pg{67--94}.

\bibitem[Vreman \& Kuerten(2014)]{Vreman14}
{\sc \au{Vreman, A.~W.} \& \au{Kuerten, J. G.~M.}} \yr{2014}  \at{Statistics of
  spatial derivatives of velocity and pressure in turbulent channel flows}.
  \jt{Phys. Fluids}  \bvol{26},  \pg{085103}.

\bibitem[Wang {\em et~al.\/}(2012)Wang, Shi, Wang, Xiao, He \&
  Chen]{Wang2012compressiso}
{\sc \au{Wang, J.}, \au{Shi, Y.}, \au{Wang, L.-P.}, \au{Xiao, Z.}, \au{He,
  X.~T.} \& \au{Chen, S.}} \yr{2012}  \at{Effect of compressibility on the
  small-scale structures in isotropic turbulence}.  \jt{J. Fluid Mech.}
  \bvol{713},  \pg{588--631}.

\bibitem[Wang {\em et~al.\/}(2018)Wang, Wan, Chen, Xie \&
  Chen]{Wang2018shockwaves}
{\sc \au{Wang, J.}, \au{Wan, M.}, \au{Chen, S.}, \au{Xie, C.} \& \au{Chen, S.}}
  \yr{2018}  \at{Effect of shock waves on the statistics and scaling in
  compressible isotropic turbulence}.  \jt{Phys. Rev. E}  \bvol{97},
  \pg{043108}.

\bibitem[Yang {\em et~al.\/}(2020)Yang, Pumir \& Xu]{Yang-2020}
{\sc \au{Yang, P.-F.}, \au{Pumir, A.} \& \au{Xu, H.}} \yr{2020}  \at{Dynamics
  and invariants of the perceived velocity gradient tensor in homogeneous and
  isotropic turbulence}.  \jt{J. Fluid Mech.}  \bvol{897},  \pg{A9}.

\end{thebibliography}

\end{document}